\documentclass[conference]{IEEEtran}
\IEEEoverridecommandlockouts
\newcommand{\ignore}[1]{} 

\newcommand{\sofie}[1]{\textbf{\textcolor{magenta}{#1}}}

\usepackage{amsmath,amssymb,amsfonts}
\usepackage{algorithmic}
\usepackage{graphicx}
\usepackage{textcomp}
\usepackage{xcolor}

\usepackage[hidelinks]{hyperref}
\usepackage{doi}

\DeclareMathOperator*{\argmin}{argmin} 

\usepackage{amsmath}
\begin{document}

\title{Location-Informed Interference Suppression Precoding Methods for Distributed Massive MIMO Systems
}

\author{\IEEEauthorblockN{1\textsuperscript{st} Emiel Vanspranghels}
\IEEEauthorblockA{\textit{Department of Electrical Engineering (ESAT)} \\
\textit{KU Leuven}\\
Leuven, Belgium \\
emiel.vanspranghels@kuleuven.be}
\and
\IEEEauthorblockN{2\textsuperscript{nd} Raquel Marina Noguera Oishi}
\IEEEauthorblockA{\textit{Department of Electrical Engineering (ESAT)} \\
\textit{KU Leuven}\\
Leuven, Belgium \\
raquelmarina.nogueraoishi@kuleuven.be}
\and
\IEEEauthorblockN{3\textsuperscript{rd} Franco Minucci}
\IEEEauthorblockA{\textit{Department of Electrical Engineering (ESAT)} \\
\textit{KU Leuven}\\
Leuven, Belgium \\
franco.minucci@ingframin.eu}
\and
\IEEEauthorblockN{4\textsuperscript{th} Sofie Pollin}
\IEEEauthorblockA{\textit{Department of Electrical Engineering (ESAT)} \\
\textit{KU Leuven}\\
Leuven, Belgium \\
sofie.pollin@kuleuven.be}
}

\maketitle

\begin{abstract}
The evolution of mobile networks towards user-centric cell-free distributed Massive MIMO configurations requires the development of novel signal processing techniques. More specifically, digital precoding algorithms have to be designed or adopted to enable distributed operation. Future deployments are expected to improve coexistence between cellular generations, and between mobile networks and incumbent services such as radar. In dense cell-free deployments, it might also not be possible to have full channel state information for all users at all antennas. To leverage location information in a dense deployment area, we suggest and investigate several algorithmic alterations on existing precoding methods, aimed at location-informed interference suppression, for usage in existing and emerging systems where user locations are known. The proposed algorithms are derived using a theoretical channel model and validated and numerically evaluated using an empirical dataset containing channel measurements from an indoor distributed Massive MIMO testbed. When dealing with measured CSI, the impact of the hardware, in addition to the location-based channel, needs to be compensated for. We propose a method to calibrate the hardware and achieve measurement-based evaluation of our location-based interference suppression algorithms.
The results demonstrate that the proposed methods allow location-based interference suppression without explicit CSI knowledge at the transmitter
, under certain realistic network conditions.
\end{abstract}


\section{Introduction}
For future mobile networks, i.e., the sixth generation (6G), a shift towards distributed Massive Multiple-Input Multiple-Output (MIMO) is anticipated, to enable user-centric cell-free network operation, and combine the benefits of Massive MIMO and network densification \cite{10054381, CHEN2022695}. Through meticulous coordination and advanced signal processing techniques, spatially distributed Access Points (APs) function as a coherent system, providing seamless connectivity and uniform quality of service. By eliminating fixed cell boundaries, cell-free networks address many inherent limitations of traditional cellular networks and offer a promising framework for enhancing spectral efficiency, coverage, and network reliability in future wireless systems \cite{9336188}. Signal processing techniques developed for cell-free networks typically require Channel State Information (CSI) from each User Equipment (UE) antenna to all AP antennas, for all APs that serve the UE.

Complementary to the cell-free paradigm, beamforming constitutes another key component of dense multi-user networks. Beamforming is a signal processing technique employed in multi-antenna systems to perform spatial filtering or steering of wireless signals, based on directional information in the far field, or location information in the near field. Signals originating from different antennas can be modified, e.g., digitally, in order to tailor the spatial power distribution, when considering transmit beamforming, or the spatial sensitivity, in the case of receive beamforming. In transmit beamforming, signals can be premultiplied to achieve constructive interference at an intended receiver, thereby concentrating transmit power and enhancing the Signal-to-Noise Ratio (SNR) relative to unfocused transmission. Such techniques do not require CSI knowledge, as steering vectors are computed from location information. In this paper we will study both far-field, or delay-and-sum beamforming, as well as near-field beamforming, or beamfocusing, which is considered an emerging technique, especially promising for future wireless communication networks \cite{10144712, 10068140}.

Another option is to utilize linear precoding techniques that make use of CSI, such as Maximum Ratio Transmission (MRT) \cite{795811}. In multi-user systems, however, it can prove beneficial to not only maximize signal power at the intended receiver, but to suppress interference at unintended receivers as well. Existing interference suppressing linear precoding techniques consist of Zero-Forcing (ZF) \cite{4599181, 1261332}, and Regularized Zero-Forcing (RZF) \cite{8668481, 1391204}.

In this paper, we propose precoding methods aimed at interference suppression, leveraging location information instead of CSI. An encountered problem is that the channel between two antennas does not only depend on the wireless medium in between, but also on hardware contributions. We propose a method to compensate for the hardware-induced phase offsets, and evaluate the location-informed interference suppressing linear precoding techniques on a calibrated channel measurement dataset. The main contributions of this paper are as follows.
\ignore{
Several mature, well-established beamforming algorithms include far-field, or delay-and-sum beamforming, Maximum Ratio Transmission (MRT) \cite{795811}, Zero-Forcing (ZF) \cite{4599181, 1261332}, and Regularised Zero-Forcing (RZF) \cite{8668481, 1391204}. Near-field beamforming, or beamfocusing, is considered an emerging technique, especially promising for future wireless communication networks \cite{10144712, 10068140}. 
\sofie{I would write: Many linear algorithms used often in the context of cell-free, could also rely on location information. E.g., knowing a location, and assuming a simple Line-of-Sight pathloss model, CSI can be estimated and used for the CSI-based cell-free processing. The only problem is that in reality, CSI does not only contain information about the propagation, but also the hardware. In this paper, we evaluate how good these methods can work for location-based interference suppression with and without hardware effects.  }

In this paper, we consider beamforming algorithms leveraging location information for the purpose of suppressing interference at unintended receivers. The main contributions of this paper are as follows.}
\begin{itemize}
    \item We propose a vector orthogonalization-based approach to generalizing zero-forcing precoding, to enable location-informed interference suppression. By reformulating the mathematics behind zero-forcing, hybrid interference suppression algorithms, exploiting both location information and channel state information, can be defined.
    \item We propose a phase calibration method for empirical channel measurement data, to ensure correct evaluation of location-informed precoding methods.
    \item Through simulations based on an empirical channel state information dataset, we demonstrate that the proposed methods constitute viable alternatives and extensions to existing precoding techniques, under certain network conditions.
\end{itemize}

The rest of the paper is organized as follows. Section \ref{sysmod} discusses the system model considered in this paper. The proposed precoding scheme is presented in Section \ref{beamalg}, followed by a description of the phase calibration procedure in Section \ref{phacal}. Section \ref{simres} provides a comprehensive presentation and discussion of the simulation results. The paper is concluded in Section \ref{conc}.

\textit{Notation:} 
$\mathbf{x}\in\mathbb{C}^{M}$ represents an $M \times 1$ complex vector, and $\mathbf{X}\in\mathbb{C}^{M\times N}$ denotes an $M\times N$ complex matrix. $\mathbf{x}_k$ is the $k$-th column of matrix $\mathbf{X}$, while $x_k$ is used for the $k$-th element of vector $\mathbf{x}$, unless stated otherwise. Element $(i,j)$ of matrix $\mathbf{X}$ is written as $\left[\mathbf{X}\right]_{i,j}$, and $\mathbf{x}^H$ and $\mathbf{X}^H$ denote the Hermitian, or complex conjugate transpose, of vector $\mathbf{x}$ and matrix $\mathbf{X}$, respectively. The complex conjugate of both is written as $\mathbf{x}^*$ and $\mathbf{X}^*$. The Euclidean norm of vector $\mathbf{x}$ is denoted by $\left \lVert \mathbf{x} \right \rVert$, and the absolute value or magnitude of a complex number $z$ is represented by $\lvert z \rvert$. The phase of a complex number $z$ is denoted by $\angle (z)$. The inverse of matrix $\mathbf{X}$ is written as $\mathbf{X}^{-1}$. $j=\sqrt{-1}$ is the imaginary unit. $\mathbb{I}_K$ denotes the identity matrix of size $K\times K$. A variable $x$ following a circularly symmetric complex Gaussian distribution with mean $\mu$ and variance $\sigma^2$, is represented by $x \sim \mathcal{CN}(\mu, \sigma^2)$. Symbols are included as lowercase letters or Greek lowercase letters, e.g. $s$ and $\alpha$, while constants can be lower- or uppercase, such as $d$ and $K$, with the exception of the constant $\pi$.

\section{System Model}\label{sysmod}
Consider an AP or Base Station (BS) consisting of $M$ transmit antennas, and single-antenna UEs. The AP transmits a symbol $s_k \in \mathbb{C}$ towards UE $k$, after precoding it with precoding vector $\mathbf{w}_k \in \mathbb{C}^{M\times 1}$. Following wireless propagation and the inclusion of hardware-induced effects, represented by the channel vector $\mathbf{h}_k\in\mathbb{C}^{M \times 1}$ towards user $k$, the digital signal processing unit at the UE receives $y_k=\mathbf{h}_k^H\mathbf{w}_ks_k+n_k$, where $n_k$ denotes circularly symmetric complex Gaussian noise with zero mean and variance $\sigma_n^2$, i.e. $n_k \sim \mathcal{CN}(0, \sigma_n^2)$. When unit symbol power is assumed, i.e. $\mathbb{E}[|s_k|^2]=1$, the SNR experienced by UE $k$ is given by $\left| \mathbf{h}_k^H\mathbf{w}_k \right|^2/\sigma_n^2$.

Furthermore, when generalizing to a multi-user MIMO system, consisting of $K$ different UEs, we define the channel matrix as the collection of channel vectors towards all users: 
\begin{equation}
\mathbf{H}=[\mathbf{h}_1, \mathbf{h}_2, \ldots,\mathbf{h}_K]\in\mathbb{C}^{M\times K}.
\label{eq:Hmatrix}
\end{equation}
The Signal-to-Interference-plus-Noise Ratio (SINR) at user $k$ is then computed as 
\begin{equation}
\mathrm{SINR}_k=\frac{|\mathbf{h}_k^H \mathbf{w}_k|^2}{\sum\limits_{\substack{l = 1 \\ l \neq k}}^K|\mathbf{h}_k^H \mathbf{w}_l|^2 + \sigma^2_n},
\end{equation}
where $|\mathbf{h}_k^H \mathbf{w}_l|^2$ is the interference experienced by UE $k$ as a result of the signal intended for UE $l$. Moreover, we normalize per-user power allocation as $||\mathbf{w}_k||^2=1$, and don't consider power allocation optimization methods.

\section{Precoding Methods}\label{beamalg}
\subsection{Existing Methods}
First, we introduce the mathematics behind the beamforming algorithms and precoding methods mentioned in the introduction, as they will be extended on in our proposed methods. For far-field beamforming, the beamforming weights are defined as 
\begin{equation}
w_i=e^{-j 2 \pi \frac{d_i \textrm{sin}(\theta)}{\lambda}},
\end{equation}
where $d_i$ denotes the distance between antenna $i$ and a reference antenna, $\theta$ is the steering angle relative to the array broadside (perpendicular to the array), and $\lambda$ is the wavelength of the carrier signal. For near-field beamforming, these weights become 
\begin{equation}
w_i=e^{-j 2 \pi \frac{d_i }{\lambda}},
\end{equation}
with $d_i$ now equal to the distance between antenna $i$ and the intended receiver. MRT deploys a precoding vector 
\begin{equation}
\mathbf{w}_k=\frac{\mathbf{h}_k}{\left \lVert \mathbf{h}_k \right \rVert},
\end{equation}
such that $|\mathbf{h}_k^H\mathbf{w}_k|$ is maximized. For zero-forcing, the precoding vector towards UE $k$ is defined as the $k$-th column of the matrix \begin{equation}
\mathbf{W}_{\textrm{ZF}}=\mathbf{H}(\mathbf{H}^{H}\mathbf{H})^{-1},
\end{equation}
such that the joint inner product of the channel vectors and the precoding vectors towards all users becomes \begin{equation}
\mathbf{H}^H\mathbf{W}_{\textrm{ZF}}=\mathbf{H}^H\mathbf{H}(\mathbf{H}^H\mathbf{H})^{-1}=\mathbb{I}_K,
\end{equation}
effectively cancelling all interference terms. By regularizing the matrix on which the inverse operation is applied, we obtain the definition of RZF: 
\begin{equation}
\mathbf{W}_{\textrm{RZF}}=\mathbf{H}(\mathbf{H}^{H}\mathbf{H} + \alpha \mathbb{I}_K)^{-1},
\end{equation}
with $\alpha$ usually, and in this work, equal to $\sigma_n^2$.

\subsection{Proposed Methods}
By rewriting the zero-forcing operation as an orthogonalization of a precoding vector with respect to a subspace, a more flexible approach to suppressing interference is enabled. This is mathematically represented as 
\begin{equation}
\mathbf{w}'=\mathbf{w} - \mathbf{V}(\mathbf{V}^H\mathbf{V})^{-1}\mathbf{V}^H\mathbf{w},
\label{eq:orthogonalisation}
\end{equation}
where $\mathbf{w}$ can be any precoding vector, and $\mathbf{V}$ can contain channel vectors or beamforming vectors, or a combination thereof. By inserting the MRT precoding vector towards UE $k$, i.e.
\begin{equation}
\mathbf{w}_k=\frac{\mathbf{h}_k}{\left \lVert \mathbf{h}_k \right \rVert},
\end{equation}
as the precoding vector $\mathbf{w}$, and defining $\mathbf{V}$ as 
\begin{equation}
\mathbf{V}=\mathbf{H}'_k=[\mathbf{h}_1, \mathbf{h}_2, \ldots,\mathbf{h}_{k-1}, \mathbf{h}_{k+1},\ldots,\mathbf{h}_K]\in\mathbb{C}^{M\times (K-1)},
\end{equation}
the resulting precoding vector, after normalization, i.e. 
\begin{equation}
\hat{\mathbf{w}}=\frac{\mathbf{w}'}{\left \lVert \mathbf{w}' \right \rVert},
\end{equation}
is equal to the normalized $k$-th column of the zero-forcing precoding matrix $\mathbf{W}_{\textrm{ZF}}$, validating the approach. The regularized variant of (\ref{eq:orthogonalisation}) is written as 
\begin{equation}
    \mathbf{w}=\mathbf{w} - \mathbf{V}(\mathbf{V}^H\mathbf{V} + \alpha \mathbb{I}_K)^{-1}\mathbf{V}^H\mathbf{w}.
    \label{eq:orthogonalisation2}
\end{equation}

The following notation is used for concrete instantiations of (\ref{eq:orthogonalisation}) and (\ref{eq:orthogonalisation2}). A precoding vector of type \texttt{a} that is orthogonalized against a subspace spanned by vectors of type \texttt{b}, is denoted by \texttt{a\_b}. MRT\_nf, for instance, is a maximum ratio transmission precoding vector towards an intended user, orthogonalized against near-field beamforming vectors towards unintended users. Depending on the used precoding vectors, CSI or location information of intended and unintended UEs is required. These algorithm-dependent requirements are included in Table \ref{tab:algorithm_comparison}. Furthermore, DIS is used to indicate distributed operation, while R stands for the use of the regularized equation. A special case is the (R)ZF\_nf algorithms, which are derived from an MRT precoding vector towards an intended user, and a combination of MRT precoding vectors and near-field beamforming vectors towards unintended users, depending on whether the unintended users are served by the same APs, and thus whether CSI is available. 

\begin{table}[!t]
    \caption{CSI and location information requirements of the different precoding algorithms.}
    \begin{center}
    \centering
    \begin{tabular}{|l|p{1.2cm}|p{1.2cm}|p{1.2cm}|p{1.2cm}|}
        \hline
        \textbf{Algorithm} & \textbf{Location intended UE\strut} & \textbf{CSI intended UE} & \textbf{Location unintended UEs\strut} & \textbf{CSI unintended UEs} \\
        \hline
        Near-field & $\checkmark$ & & & \\
        \hline
        MRT & & $\checkmark$ & & \\
        \hline
        (DIS) (R)ZF & & $\checkmark$ & & $\checkmark$ \\
        \hline
        nf\_nf & $\checkmark$ & & $\checkmark$ & \\
        \hline
        (DIS) (R)MRT\_nf & & $\checkmark$ & $\checkmark$ & \\
        \hline
        (R)ZF\_nf & & $\checkmark$ & $\checkmark$$^a$ & $\checkmark$$^a$ \\
        \hline
        \multicolumn{5}{p{8.5cm}}{\footnotesize $^{\mathrm{a}}$Location information for inter-cluster interference suppression, CSI for intra-cluster interference suppression.}

    \end{tabular}
    \label{tab:algorithm_comparison}
    \end{center}
\end{table}

\section{Phase Calibration}\label{phacal}
The location-based algorithms depend on two crucial elements for their performance to be accurately assessed. The first is a predominantly Line-of-Sight (LoS) wireless channel. For this purpose, a dense dataset containing CSI measurements from an indoor distributed LoS setup, collected through the KU Leuven Massive MIMO testbed, is utilized\footnote{Available: https://dx.doi.org/10.21227/nr6k-8r78.} \cite{nr6k-8r78-21} \cite{9535488}. The second requirement relates to the elimination of hardware-induced contributions to the channel state information. If the antennas of a multi-antenna system are not calibrated to share a common phase reference, location-based precoding methods that do not take into account these inter-antenna variations will not be effective, as opposed to CSI-informed techniques, which inherently consider the hardware contributions to the CSI. 
\ignore{\sofie{I think it is not exactly as you say. First of all, if the phase imbalance is stable (coherent) over the block, then I don't think it is a problem and all these effects are compensated in the full CSI. That is how systems work, you just estimate it and assume it will be stable/coherent. The problem I think is that when you do location-based beamforming, you need also a way to sync the phases of the hardware, or estimate them, or compensate for them in addition to compensating for the location. }
Agreed, MRT for example perfectly compensates for this by using the FULL CSI, including hardware contributions. I added 'location-based' to specify it is only algorithms that do not use CSI that suffer from these phase offsets.}

It is important to note that the phase offsets introduced by the hardware depend on the Transmit (Tx) and Receive (Rx) antennas $t_k$ and $r_i$, and it is necessary to estimate a joint phase compensation factor $\phi_{t_k}^{r_i}$ per antenna pair, and thus per antenna-to-antenna channel. We can estimate these phase offsets in the following way. \ignore{For this purpose, the channel state information values in the dataset are phase-calibrated in the following way.} For each Tx and Rx antenna pair, a phase offset compensation factor was computed by minimizing the Mean Squared Error (MSE) between the phase of the theoretical LoS channel and the compensated measurement data, as depicted by

\setlength{\arraycolsep}{0.0em}
\begin{align}
\varphi_{t_k}^{r_i} &= \argmin_{\varphi} \Bigg( \sum_m \sum_n \Big| \exp\left(j \angle\left( [\mathbf{H}_{t_k, r_i}^{\textrm{LoS}}]_{m,n} \right) \right) \nonumber \\
&\quad {} - \exp(j \varphi) \cdot \exp\left(j \angle\left( [\mathbf{H}_{t_k, r_i}^{\textrm{emp}}]_{m,n} \right) \right) \Big|^2 \Bigg).
\label{eq:argmin}
\end{align}
\setlength{\arraycolsep}{5pt}
The theoretical LoS channel, which underpins both far-field and near-field beamforming, describes the propagation of a wireless signal with wavelength $\lambda$ over a path of length $d$, resulting in a phase offset $\varphi = -\frac{2\pi d}{\lambda}$.

The solution to (\ref{eq:argmin}) is computed as 
\setlength{\arraycolsep}{0.0em}
\begin{align}
\varphi_{t_k}^{r_i}&=\angle \Bigg( \sum_{m} \sum_{n}\big( \exp(j\angle(\left[\mathbf{H}_{t_k, r_i}^{\textrm{LoS}}\right]_{m,n})) \cdot \nonumber \\ & \exp(j\angle(\left[\mathbf{H}_{t_k, r_i}^{\textrm{emp}}\right]_{m,n}^*))\big)\Bigg),
\label{eq:correlation2}
\end{align}
\setlength{\arraycolsep}{5pt}

where the sum is taken over all available data points in the dataset originating from transmit antenna $t_k \in (1,2,3,4)$ and receive antenna $r_i \in (1,2,\ldots,64)$. These matrices $\mathbf{H}$ do not comply with (\ref{eq:Hmatrix}), but contain CSI values between two specific antennas, over an $xy$-grid of measurement positions. 

\begin{figure}[htbp]
    \centering
    \includegraphics[width=\linewidth]{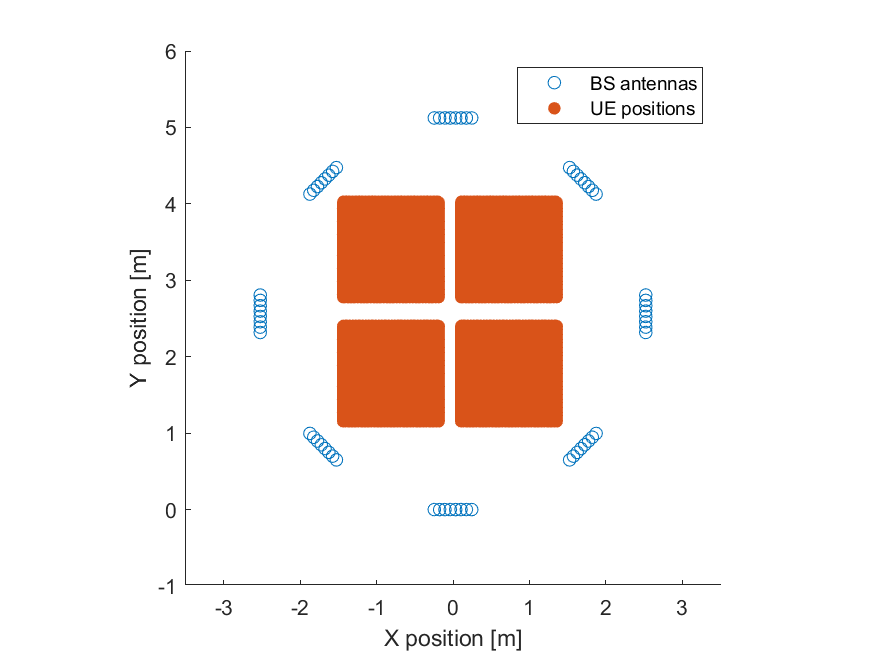}
    \caption{Base station antenna locations and user equipment antenna measurement positions.}
    \label{fig:setup}
\end{figure}

Fig. \ref{fig:setup} shows the measurement setup used to acquire the CSI dataset, in which data for the four distinct $xy$-grids were obtained by four antennas $t_k$ transmitting uplink pilot signals, received by the 64 $r_i$ antennas surrounding the Region-of-Interest (RoI). Therefore, the UE antennas are denoted as the Tx antennas $t_k$, and the BS antennas as the Rx antennas $r_i$, and we leverage the concept of physical channel reciprocity within a coherence time in Time Division Duplex (TDD) systems, after removing the hardware contributions from the channel data, in order to evaluate downlink precoding.

\ignore{\sofie{It is important to note that this hardware calibration does not depend on the location of the incumbent, and can be carried out easily in a system where there are many transmit and receive antennas with know location and precise timestamping performance. } The location is needed to compute the theoretical LoS phase offset }

\begin{figure}[htbp]
    \centering
    \includegraphics[width=0.7\linewidth]{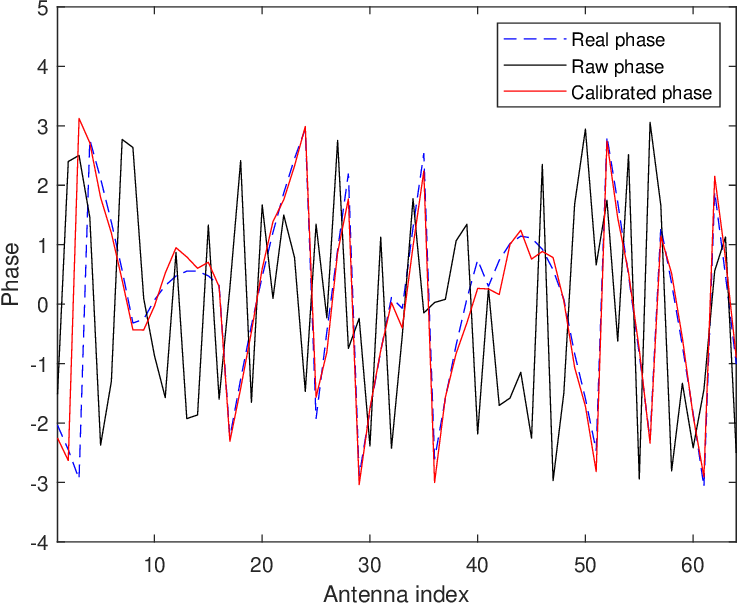}
    \caption{The phase of the raw CSI, the calibrated CSI and the theoretical LoS channel model.}
    \label{fig:phase_figure4cd}
\end{figure}

\ignore{
The measurements were conducted using uplink pilot sequences, which is why the UE antennas are denoted as transmit antennas, and the BS antennas as receive antennas. We therefore leverage the concept of physical channel reciprocity within a coherence time in Time Division Duplex (TDD) systems, after removing the hardware contributions from the channel data, in order to evaluate downlink precoding.} The result of the phase calibration procedure is illustrated in Fig. \ref{fig:phase_figure4cd}, showing a phase comparison between the raw CSI, the calibrated CSI and the theoretical LoS channel, between one of the UE measurement positions and the 64 BS antennas.

\section{Simulation Results}\label{simres}
In this section, simulation results from two multi-user scenarios are presented and discussed. The 64-antenna base station, shown in Fig. \ref{fig:setup}, used during the measurements, is virtually split into eight access points consisting of an antenna array of eight antennas each. In the first scenario, all users are served by all access points simultaneously, either with centralized or distributed signal processing. In the second scenario, the users are clustered into groups, each served by a predefined pair of access points, assigned based on maximum average channel gain. Because far-field beamforming-based algorithms did not prove competitive in the distributed array case, they are left out in this section.

\subsection{Full Access Point Coordination}
\begin{figure}[htbp]
    \centering
    \includegraphics[width=0.75\linewidth]{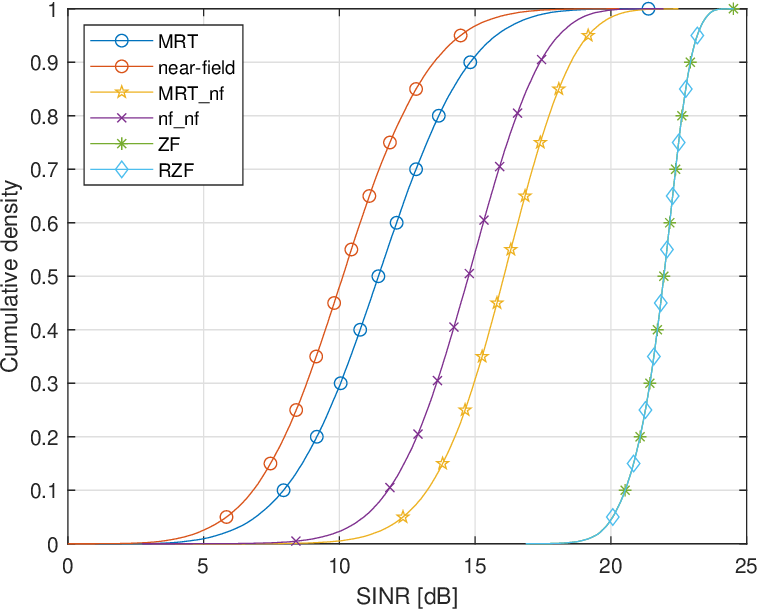}
    \caption{CDF of the SINR of different precoding methods.}
    \label{fig:cdf_multi_4_paper}
\end{figure}

The results in Fig. \ref{fig:cdf_multi_4_paper} were obtained from Monte Carlo simulations using the channel measurement dataset, considering $K=5$ UEs, uniformly distributed over the $xy$-grid, spaced at least 10cm from each other, and a noise floor of -20dB relative to the average received power. The performance of near-field beamforming is shown to approach that of MRT, validating the LoS-predominance and phase calibration procedure. MRT\_nf and nf\_nf are found to outperform MRT, but logically fall short of ZF and RZF, which require CSI towards all users.

\ignore{
\sofie{How did you compute SINR, is that equation 2? } Yes

\sofie{You have 5 users, and if I understand correctly, in MRT you only do precoding based on the known CSI of the intended user. IN ZF you have CSI of all users? And in nfnf you have CSI of no users? Only location info? And in MRTnf you have CSI of 1 user and location of the other users? Make that more explicit}
}

\begin{figure}[htbp]
    \centering
    \includegraphics[width=0.75\linewidth]{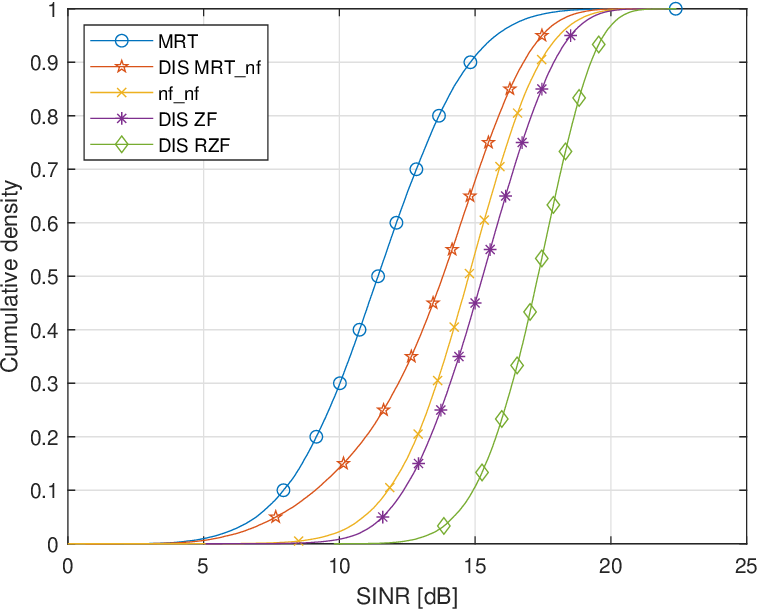}
    \caption{CDF of the SINR of different precoding methods, with fully distributed operation.}
    \label{fig:cdf_multi_dis}
\end{figure}

Fig. \ref{fig:cdf_multi_dis} shows the same results for distributed operation, i.e., all UEs are still served by all APs, but the signal processing is performed locally at the APs, instead of in a centralized manner. Network-wide location information is assumed to be available at all APs, while each AP only has access to its own CSI towards the UEs, limiting the orthogonalization procedure to a lower-dimensional vector space. nf\_nf does not suffer from this restriction, as it does not leverage CSI, and proves to perform almost as well as ZF. RZF significantly outperforms ZF in this scenario because of the orthogonalization procedure in a lower-dimensional vector space, with a smaller amount of degrees of freedom, yielding a reduced residual component of the original precoding vector, and thus a smaller received signal power at the intended user, and a larger relative influence of noise on the SINR. RZF essentially trades off some interference suppression at unintended receivers in favor of improved noise suppression at the intended receiver.

\begin{figure}[htbp]
    \centering
    \includegraphics[width=0.75\linewidth]{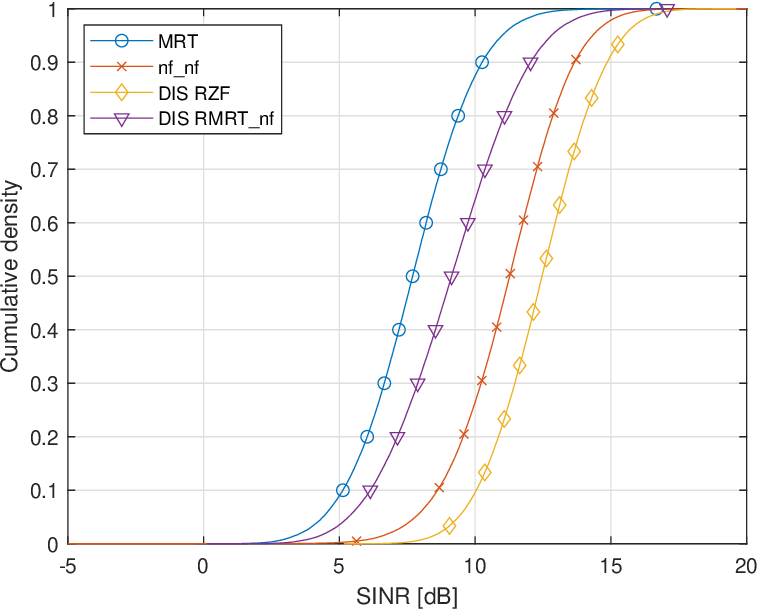}
    \caption{CDF of the SINR of different precoding methods, with fully distributed operation, and 10 UEs.}
    \label{fig:cdf_multi_dis_10}
\end{figure}

Even more promising results are obtained by letting $K=10 > M=8$, i.e., a higher number of UEs than transmit antennas at individual APs. This forces the matrix $\mathbf{V}$ from (\ref{eq:orthogonalisation}) to not have full column rank, and subsequently, regularization of $\mathbf{V}^H\mathbf{V}$ is required. Fig. \ref{fig:cdf_multi_dis_10} illustrates that as the number of UEs grows larger, i.e. $K=10$, especially beyond the number of antennas per AP, the performance of nf\_nf increases relative to that of RZF.

\begin{figure}[htbp]
    \centering
    \includegraphics[width=0.75\linewidth]{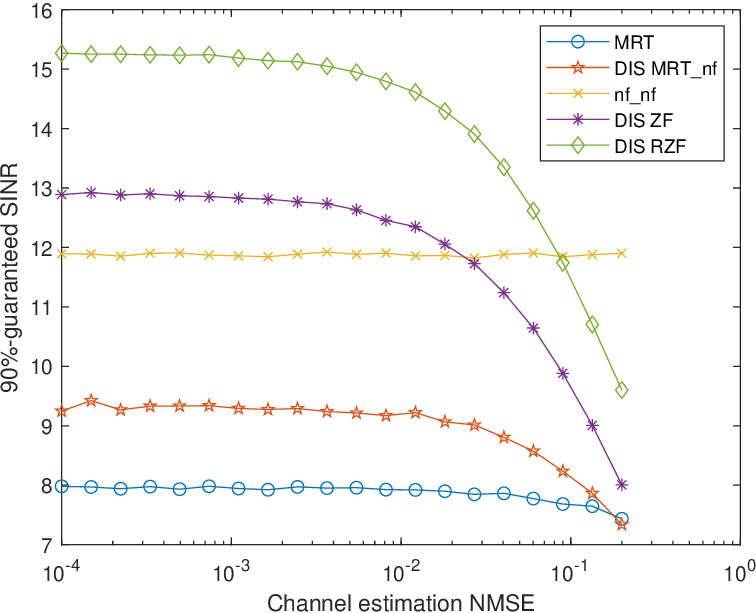}
    \caption{90\%-guaranteed SINR of different precoding methods in function of relative channel estimation errors, with fully distributed operation.}
    \label{fig:quantile_sinr2}
\end{figure}

\begin{figure}[htbp]
    \centering
    \includegraphics[width=0.75\linewidth]{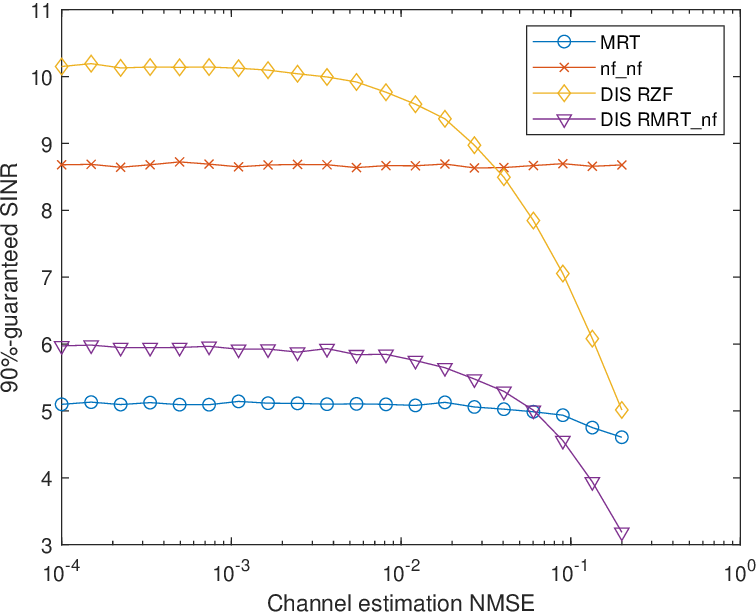}
    \caption{90\%-guaranteed SINR of different precoding methods in function of relative channel estimation errors, with fully distributed operation, and 10 UEs.}
    \label{fig:quantile_sinr3}
\end{figure}

On top of the distributed operation, the performance of the algorithms was evaluated under the presence of channel estimation errors, both for the case where $K=5<M=8$ and $K=10>M=8$. The results are shown in Fig. \ref{fig:quantile_sinr2} and Fig. \ref{fig:quantile_sinr3}, now as the SINR that is guaranteed for 90\% of the UE position combinations, in function of the Normalized Mean Squared Error (NMSE) of the channel estimates. The channel estimation errors are modelled as i.i.d. circularly symmetric complex Gaussian variables, i.e. $\left[\mathbf{H}^{\textrm{err}}\right]_{i,j}\sim \mathcal{CN}(0, \sigma_E^2)$, so that the channel estimates are equal to $\mathbf{\hat{H}}=\mathbf{H}+\mathbf{H}^{\textrm{err}}$. The NMSE is then computed as $\frac{1}{IJ}\sum_{i,j}|\left[\mathbf{H}^{\textrm{err}}\right]_{i,j}|^2$ divided by $\frac{1}{IJ}\sum_{i,j}|\left[\mathbf{H}\right]_{i,j}|^2$.

We can clearly observe that MRT suffers less from channel estimation errors than ZF approaches; it is particularly important for channel estimates towards unintended users to be accurate. Therefore, in the absence of precise channel estimation procedures, leveraging location information can be advantageous for interference suppression or even for constructing the entire precoding vector, as done by nf\_nf. However, we notice that MRT\_nf and its regularized variant also experience a significant performance degradation due to channel estimation errors, despite not relying on channel estimates for the purpose of interference suppression. This loss of performance is more severe than for MRT itself, because MRT does not require vector orthogonalization. When orthogonalizing against multiple precoding vectors in a low-dimensional vector space, the accuracy of the channel estimate prior to orthogonalization becomes more critical, as a significant part of the amplitude pointing in the correct direction can get removed.

\subsection{Network-Centric Clustering}
\begin{figure}[htbp]
    \centering
    \includegraphics[width=\linewidth]{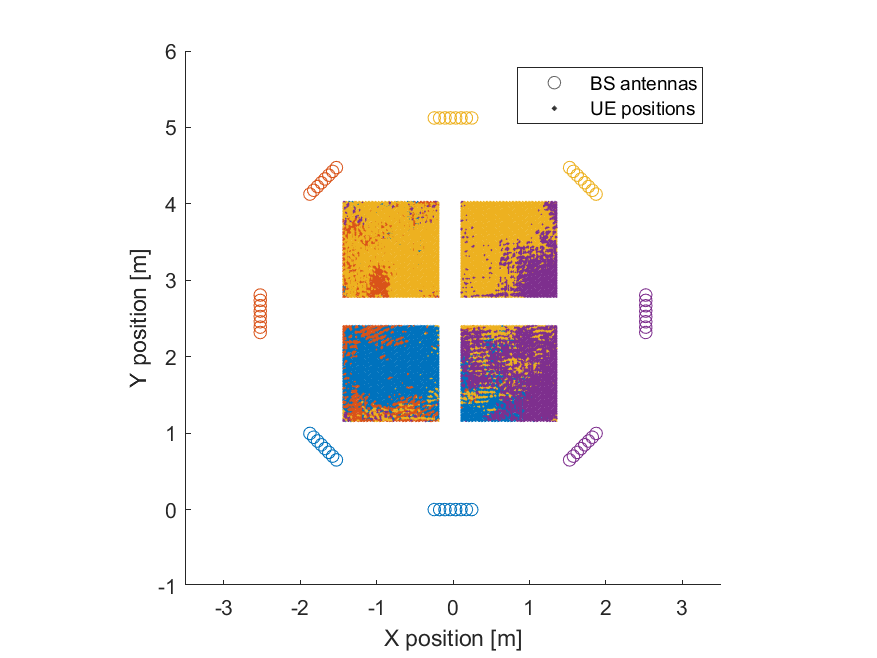}
    \caption{Clustering based on channel gain.}
    \label{fig:cluster_gain}
\end{figure}

This experiment aligns more closely with envisioned scalable user-centric cell-free networks, in which UEs are served only by a subset of the available APs. For simplicity, we applied network-centric clustering, where the 64 antennas are split up into predefined pairs of two APs, each pair having access to two times eight antennas and a shared processing unit. The UEs are assigned two access points, i.e. one of the four pairs, based on average channel gain towards the antennas of the pairs. The result of clustering all UE positions in that way is shown in Fig. \ref{fig:cluster_gain}.

\begin{figure}[htbp]
    \centering
    \includegraphics[width=0.75\linewidth]{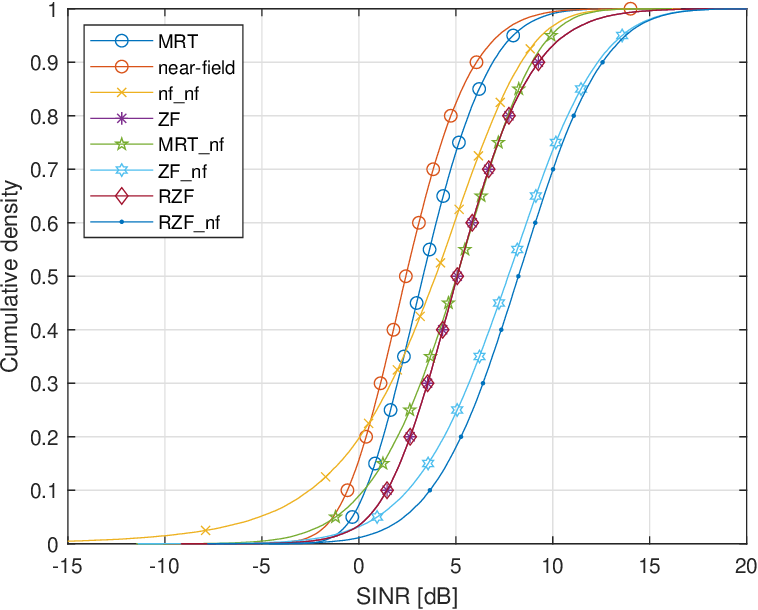}
    \caption{CDF of the SINR of different precoding methods, in the clustering-based multi-user scenario, with 10 UEs.}
    \label{fig:cdf_clustering}
\end{figure}

This setup provides insight into a valuable enhancement of existing algorithms. Suppose that pairs of APs have access to the CSI of the UEs they serve, supplemented by location information of other users in the network. This way, intra-cluster interference may be suppressed using CSI, while inter-cluster interference can be suppressed using location information. We then come to a (R)ZF\_nf approach, where CSI and near-field beamforming vectors are combined in matrix $\mathbf{V}$ from (\ref{eq:orthogonalisation}) and (\ref{eq:orthogonalisation2}). The results, included in Fig. \ref{fig:cdf_clustering}, are promising, as ZF\_nf and RZF\_nf effectively increase the 90\%-guaranteed SINR relative to all other algorithms, while the enhancement is even greater for UEs experiencing better network conditions, increasing the achievable peak data rates.

\section{Conclusion}\label{conc}

In this work, we investigated the performance of well-known precoding methods in a distributed Massive MIMO setting. We proposed a zero-forcing generalization to allow for beamforming vector-based interference suppression, enabling the derivation of precoding vectors leveraging both CSI and location information, with emphasis on location-informed interference suppression. Using empirical channel measurements, we have shown that this strategy provides viable alternatives and extensions to existing precoding methods. We proposed a low-complexity method for the phase calibration of empirical CSI data, to allow for a reliable evaluation of location-informed precoding methods.

We hope this work inspires future research towards (location-informed) channel prediction-based interference suppression, and signal processing techniques for dense cell-free wireless networks in general. Specific suggestions are the implementation of the algorithms in a distributed Massive MIMO testbed, as well as simulations considering user-centric clustering, in which hybrid vectors consisting of both CSI values and location-based beamforming coefficients may be used to compute the orthogonalized precoding vectors.

\ignore{
\section{TO-DO}
- American English, -ized, -ization, -zing \\
- Abbreviations only at first appearance
}

\ignore{\section*{Acknowledgment}}
\bibliography{references}

@ARTICLE{10054381,
  author={Wang, Cheng-Xiang and You, Xiaohu and Gao, Xiqi and Zhu, Xiuming and Li, Zixin and Zhang, Chuan and Wang, Haiming and Huang, Yongming and Chen, Yunfei and Haas, Harald and Thompson, John S. and Larsson, Erik G. and Renzo, Marco Di and Tong, Wen and Zhu, Peiying and Shen, Xuemin and Poor, H. Vincent and Hanzo, Lajos},
  journal={IEEE Communications Surveys \& Tutorials}, 
  title={On the Road to 6G: Visions, Requirements, Key Technologies, and Testbeds}, 
  year={2023},
  volume={25},
  number={2},
  pages={905-974},
  doi={10.1109/COMST.2023.3249835}}

@article{CHEN2022695,
    title = {A survey on user-centric cell-free massive MIMO systems},
    journal = {Digital Communications and Networks},
    volume = {8},
    number = {5},
    pages = {695-719},
    year = {2022},
    issn = {2352-8648},
    doi = {https://doi.org/10.1016/j.dcan.2021.12.005},
    url = {https://www.sciencedirect.com/science/article/pii/S2352864821001024},
    author = {Shuaifei Chen and Jiayi Zhang and Jing Zhang and Emil Bj{\"o}rnson and Bo Ai}}

@BOOK{9336188,
  author={Demir, {\"o}zlem Tugfe and Bj{\"o}rnson, Emil and Sanguinetti, Luca},
  booktitle={Foundations of User-Centric Cell-Free Massive MIMO},
  title={Foundations of User-Centric Cell-Free Massive MIMO},
  publisher={arXiv},
  year={2021},
  volume={},
  number={},
  pages={},
  doi={}}

@ARTICLE{795811,
  author={Lo, T.K.Y.},
  journal={IEEE Transactions on Communications}, 
  title={Maximum ratio transmission}, 
  year={1999},
  volume={47},
  number={10},
  pages={1458-1461},
  doi={10.1109/26.795811}}

@ARTICLE{4599181,
  author={Wiesel, Ami and Eldar, Yonina C. and Shamai, Shlomo},
  journal={IEEE Transactions on Signal Processing}, 
  title={Zero-Forcing Precoding and Generalized Inverses}, 
  year={2008},
  volume={56},
  number={9},
  pages={4409-4418},
  doi={10.1109/TSP.2008.924638}}

@ARTICLE{1261332,
  author={Spencer, Q.H. and Swindlehurst, A.L. and Haardt, M.},
  journal={IEEE Transactions on Signal Processing}, 
  title={Zero-forcing methods for downlink spatial multiplexing in multiuser MIMO channels}, 
  year={2004},
  volume={52},
  number={2},
  pages={461-471},
  doi={10.1109/TSP.2003.821107}}

@ARTICLE{8668481,
  author={Nguyen, Long D. and Tuan, Hoang Duong and Duong, Trung Q. and Poor, H. Vincent},
  journal={IEEE Transactions on Signal Processing}, 
  title={Multi-User Regularized Zero-Forcing Beamforming}, 
  year={2019},
  volume={67},
  number={11},
  pages={2839-2853},
  doi={10.1109/TSP.2019.2905833}}

@ARTICLE{1391204,
  author={Peel, C.B. and Hochwald, B.M. and Swindlehurst, A.L.},
  journal={IEEE Transactions on Communications}, 
  title={A vector-perturbation technique for near-capacity multiantenna multiuser communication-part I: channel inversion and regularization}, 
  year={2005},
  volume={53},
  number={1},
  pages={195-202},
  doi={10.1109/TCOMM.2004.840638}}

@ARTICLE{10144712,
  author={Elbir, Ahmet M. and Mishra, Kumar Vijay and Vorobyov, Sergiy A. and Heath, Robert W.},
  journal={IEEE Signal Processing Magazine}, 
  title={Twenty-Five Years of Advances in Beamforming: From convex and nonconvex optimization to learning techniques}, 
  year={2023},
  volume={40},
  number={4},
  pages={118-131},
  doi={10.1109/MSP.2023.3262366}}

@ARTICLE{10068140,
  author={Zhang, Haiyang and Shlezinger, Nir and Guidi, Francesco and Dardari, Davide and Eldar, Yonina C.},
  journal={IEEE Communications Magazine}, 
  title={6G Wireless Communications: From Far-Field Beam Steering to Near-Field Beam Focusing}, 
  year={2023},
  volume={61},
  number={4},
  pages={72-77},
  doi={10.1109/MCOM.001.2200259}}

@misc{nr6k-8r78-21,
doi = {10.21227/nr6k-8r78},
url = {https://dx.doi.org/10.21227/nr6k-8r78},
author = {Sibren De Bast and Sofie Pollin},
publisher = {IEEE Dataport},
title = {Ultra Dense Indoor MaMIMO CSI Dataset},
year = {2021} }

@ARTICLE{9535488,
  author={Li, Chenglong and De Bast, Sibren and Tanghe, Emmeric and Pollin, Sofie and Joseph, Wout},
  journal={IEEE Sensors Journal}, 
  title={Toward Fine-Grained Indoor Localization Based on Massive MIMO-OFDM System: Experiment and Analysis}, 
  year={2022},
  volume={22},
  number={6},
  pages={5318-5328},
  doi={10.1109/JSEN.2021.3111986}}
\end{document}